\documentstyle[preprint,aps,epsf]{revtex}
\tightenlines
\begin{document}
\draft
\title{ Homogeneous nucleation of quark-gluon plasma, finite
size effects and long-lived metastable objects
 }
\author{ E.E. Zabrodin$^{1,2}$, L.V. Bravina$^{1,2,}$\footnote{
Alexander von Humboldt Foundation Fellow}~, 
H. St{\"o}cker$^{1}$ and W. Greiner$^{1}$}
\address{
$^1$ Institute for Theoretical Physics, University of Frankfurt,
Robert-Mayer-Str. 8-10, D-60054 Frankfurt, Germany \\
$^2$ Institute for Nuclear Physics, Moscow State University,
119899 Moscow, Russia \\
}

\date{\today}
\maketitle

\begin{abstract}
The general formalism of homogeneous nucleation theory is applied
to study the hadronization pattern of the ultra-relativistic 
quark-gluon plasma (QGP) undergoing a first order phase transition.
A coalescence model is proposed to describe the evolution dynamics of
hadronic clusters produced in the nucleation process. The size
distribution of the nucleated clusters is important for the 
description of the plasma conversion. The model is most sensitive to 
the initial conditions of the QGP thermalization, time evolution of 
the energy density, and the interfacial energy of the 
plasma$-$hadronic matter interface. The rapidly expanding QGP is first
supercooled by about $\Delta T = T - T_c = 4-6\,\%$. Then it reheats 
again up to the critical temperature $T_c$. Finally it breaks up into 
hadronic clusters and small droplets of plasma. This fast dynamics
occurs within the first $5-10\,$fm/$c$. The finite size effects and 
fluctuations near the critical temperature are studied. It is shown 
that a drop of longitudinally expanding QGP of the transverse radius 
below 4.5$\,$fm can display a long-lived metastability.
However, both in the rapid and in the delayed hadronization scenario,
the bulk pion  yield is emitted by sources as large as $3-4.5\,$fm. 
This may be detected experimentally both by a HBT interferometry 
signal and by the analysis of the rapidity distributions of particles 
in narrow $p_T$-intervals at small $|p_T|$ on an event-by-event basis.
\end{abstract}

\widetext

\section{INTRODUCTION}
\label{sec1}

The hadronization of quark$-$gluon plasma (QGP) possibly produced in
the early Universe or in ultra-relativistic heavy ion collisions has
received much activity during the last decades \cite{Shur80,CFS74,
Bjor83,Witt84,StGr86,ClSt86,KMR86,KMLGR86,GKKM84,GKS87,RFWSG90,MiPa90,
Shur92,Bjor92,CsKa92a,CsKa92b,CKKZ93,Geig95,HuSh95,RMP95,WeAi95}.
Despite the significant progress in understanding of the
variety of possible signals and features of the QGP, the nature of
the phase transition (PT) between deconfined and confined phase is 
not clear yet. Assuming a first order PT, usually an adiabatic 
scenario is invoked to describe the conversion of plasma into hadrons. 
A few years ago, in \cite{CsKa92a} it was pointed out that the 
coarse-grained field theory of homogeneous nucleation 
\cite{Lang67,Lang69} can be relevant for the relativistic PT also. 
This scenario assumes the nucleation of hadronic bubbles, e.g. bubbles 
of pion gas, within the (initially homogeneous) supercooled 
metastable QGP as the starting point of the PT. These bubbles are 
nucleated because of the thermodynamic fluctuations of the energy 
density in the system. Then bubbles with radii smaller than the 
critical radius $R_c$ shrink, bubbles of the critical size are in
metastable equilibrium, while bubbles with radii larger than $R_c$
gain in size and develop into the new phase. The treatment of
the relaxation of the metastable state within the framework of the
nucleation theory provides the fundamental
nucleation rate \cite{Lang69,CaCo77,Affl81}, which expresses 
the number of viable nucleating clusters of the new phase via the 
equilibrium number of critically large clusters.

Langer's theory has been applied \cite{CsKa92b,CKKZ93} to 
calculate the hadronization of rapidly expanding baryon-free QGP, 
produced in heavy ion collisions at RHIC and LHC energies.
Although this approach seems to be more realistic than the idealized
adiabatic PT, several questions remain open: These
calculations \cite{CKKZ93} have shown that the latent heat
released during the plasma conversion is not sufficient to prevent
the strong $(20\,\% - 35\,\%)$ supercooling of the system.
Thus, the rapidly quenched system leaves the region of metastability 
and enters the highly unstable spinodal region.
Here the theory of spinodal decomposition might be in order to 
describe the further evolution of the fluctuations leading to the
break-up of that system.

Secondly, since the critical radius drops quickly when the temperature 
is lowered, the bulk creation of the hadronic phase should begin
\cite{CsKa92b} when the bubble radii are $r \leq 0.8\,$fm. Finally, 
in the non-scaling scenario the bubbles grow independently on the 
total expanding volume. Then the completion of the PT will be 
significantly delayed \cite{CKKZ93}. Even within the scaling scenario
the time necessary for the completion of the PT varies from 50 to 
90$\,$fm/$c$ (depending strongly on the numerical value of the surface 
tension).

Hence, from the above one may conclude that either 

$-$ the homogeneous nucleation scheme is inappropriate to
describe the hadronization of relativistic systems or 

$-$ that some important features of the first order PT are still
missing. 
The situation would change if it turned out that the amount 
of plasma converted into hadrons had been underestimated in earlier 
works. 

Recently, the calculation of the dynamical factor $\kappa$ governing 
the growth rate of subcritical bubbles was reexamined \cite{RuFr96}, 
and the size distribution of bubbles in configuration space has been 
used to estimate the supercooling of rapidly expanding QGP 
\cite{ZBCSG97}. The latter plays an important role in the 
hadronization of plasma produced in relativistic collisions.
When the critical radius of hadronic bubbles drops due to the rapid
fall of the temperature in the expanding system, the subcritical 
bubbles transfer to the region of supercritical sizes for these new
conditions in the system. These bubbles then stop to shrink and start 
to gain size, thus increasing the total volume of the hadronic phase.  

In the present paper we study the effect of the bubble size 
distribution on the dynamics of the plasma$-$hadrons phase transition. 
The paper is organized as follows: the model used to study the QGP 
hadronization is described in Sec.~\ref{sec2}. Section~\ref{sec3} 
reviews the formalism of nucleation theory. The evaluation of both 
dynamical and statistical prefactors appearing in Langer's theory are
discussed. A coalescence-type model for further evolution of the 
nucleated hadronic bubbles is proposed. The role of the initial 
conditions (as well as effects of variations of the model parameters, 
i.e. the value of the surface tension, the minimum size of the pionic 
bubbles, the non-scaling regime and dilution factor, the prefactors, 
etc., on the relaxation of the metastable QGP is studied in 
Sec.~\ref{sec4}. Section~\ref{sec5} presents the investigation of the 
finite size effects, the creation of long-lived states with metastable
QGP and hadronic bubbles, and temperature fluctuations in the system. 
Finally, the results are summarized in the Conclusions.
 
\section{THE MODEL}
\label{sec2}

We consider a QGP produced in collisions of two heavy ions at RHIC or 
LHC energies. It is assumed that the plasma is thermalized soon. A 
wide range of initial conditions is studied, see Sec.\ref{sec4}. 
The expansion and cooling is ruled by relativistic hydrodynamics.

When the plasma cools below the critical temperature $T_c$, a first 
order phase transition is initiated by the appearance of hadronic 
bubbles. The bag model equation of state (EOS) for the QGP consisting 
of gluons and massless quarks reads
\begin{equation}
\displaystyle
p_q = \frac{\pi^2}{90} \left( 16 + \frac{21}{2} n_f \right) T^4 -
B = a_q T^4 - B 
\label{1}
\end{equation}
with the number of flavors, $n_f$, and the bag constant, $B$. The EOS 
of the relativistic pion gas is 
\begin{equation}
\displaystyle
p_h = \frac{\pi^2}{10} T^4 = a_h T^4 \ .
\label{2}
\end{equation}
By imposing the condition $p_q = p_h$ at $T = T_c$ one may find the 
critical temperature
\begin{equation}
\displaystyle
T_c = \left( \frac{B}{a_q - a_h} \right)^{1/4} \ .
\label{3}
\end{equation}
For a two-flavored QGP with $B^{1/4} = 235\,$MeV \cite{CsKa92a} this
gives $T_c = 169\,$MeV.

The total energy density of the mixed quark$-$hadron phase, $e$, can
be treated in the capillary (thin wall) approximation as a linear 
combination of the energy density of the hadronic phase, $e_h$, in 
the fraction of the volume occupied by hadronic bubbles, 
$h = V_h/V_{tot}$, the energy density of the QGP, $e_q$, in the rest 
of the volume $V_q = (1 - h) V_{tot}$, and the energy density
of quark$-$hadronic interface, $e_S = \sigma S_h/V_{tot}$:
\begin{equation}
\displaystyle
e = h e_h  + (1 - h) e_q + \sigma \frac{S_h}{V_{tot}}\ .
\label{4}
\end{equation}
Here $\sigma$ is the surface tension of the interface between the 
two phases.
The last term in (\ref{4}) is usually disregarded 
\cite{CsKa92a,CsKa92b,CKKZ93} though its contribution to the total
energy density may be comparable with the other two: the ratio of the
interface between the phases to the hadronic volume is not too small.
For instance, assume that all bubbles are of the same radius,
$| R | = \alpha$. Then, at $\sigma = 0.1 T_c^3$ \cite{Iwa94}
the surface energy density scales as $e_S \approx \alpha^{-1} e_h$. 
Therefore, the surface term in Eq.(\ref{4}) may be omitted only 
for sufficiently large ($r \geq 4\,$fm) hadronic clusters.
 
Similarly to the energy density, the total pressure also consists of
three terms 
\begin{equation}
\displaystyle
p = h p_h  + (1 - h) p_q + \sum c_i p_L^{(i)} \ ,
\label{5}
\end{equation}
where the pressure of the spherical surface of radius $R_i$ is given 
by the Laplace formula, $p_L^{(i)} = 2 \sigma / R_i$, and $c_i$ is the 
local concentration of bubbles of radius $R_i$ in the total volume. 

The Bjorken model \cite{ Bjor83} of scaling longitudinal expansion is 
applied to find the time evolution of the energy density. It yields 
the derivative of the energy density with respect to the proper 
time, $\tau$,
\begin{equation}
\displaystyle
\frac{{\rm d} e}{{\rm d} \tau} = - \frac{e + p}{\tau} \ .
\label{6}
\end{equation}

Next one needs to compute the fraction of the total volume converted 
to hadronic phase. The pionic bubbles, appeared because of the 
fluctuations in the energy density, will either shrink or grow. This 
problem cannot been solved on the basis of the thermodynamic theory 
of fluctuations only. It requires a kinetic description of the process 
of bubble evolution. The volume fraction $h(R,t)$ of bubbles of size 
$R$ at time $t$ obeys the equation of motion in the size space
\begin{equation}
\displaystyle
\frac{\partial h(R,t)}{\partial t} = - \frac{\partial}{\partial R}
\left[ v(R) h(R,t) \right] + h^{nucl}(R,t) \ ,
\label{7}
\end{equation}
where $v(R)$ is the radial velocity and $h^{nucl}$ denotes the hadronic 
fraction created at time $t$. Without the nucleation term Eq.(\ref{7}) 
transforms into the continuity equation of the Lifshitz-Slyozov theory
\cite{LiSl61}. The volume fraction $h^{nucl}(R)$ of bubbles of size $R$
nucleated per unit time is given \cite{BrZa96} by the distribution
\begin{eqnarray}
\displaystyle
h^{nucl}(R) &=& \frac{I}{\sqrt{2 \pi (9 \tau + 2 \lambda_Z^2)}} 
\exp \left[ \frac{9 \tau + 2 \lambda_Z^2}{2}\, (r - 1)^2 \right] \\
\label{8}
&\times & 
\int_r^\infty a \, \left[ 3 \tau (a^2 + a + 1) + 2 \lambda_Z^2 a^2
\right]\, \exp \left[ - \frac{9 \tau + 2 \lambda_Z^2}{2}\,
(a - 1)^2 \right] {\rm d}a \ ,
\nonumber
\end{eqnarray}
containing the nucleation rate $I$, the critical exponent $\tau$ and 
the two dimensionless variables $\lambda_Z$ and $r$ (see below), so 
that
\begin{equation}
\displaystyle
\int_0^\infty h^{nucl}(R)\, {\rm d}R = I\ .
\label{9}
\end{equation}
Instead of using a continuous spectrum, hadronic matter in our model 
is represented by a discrete spectrum of pion bubbles starting from
$r_0 = 1\,$fm.

\section{THEORY OF HOMOGENEOUS NUCLEATION}
\label{sec3}

The homogeneous nucleation has been subject of intensive investigation 
both theoretically and experimentally for a long period (for reviews
see \cite{Fren46,GMS83,Kelt91} and references therein). In our 
analysis we follow the coarse-grained theory by Langer 
\cite{Lang67,Lang69}, who extended the classical 
Becker-D{\"o}ring-Zeldovich (BDZ) theory of nucleation 
\cite{BeDo35,Zeld43} to field theories. The nucleation rate in both 
classical and modern coarse-grained field theory reads
\begin{equation}
\displaystyle
I = I_0\, \exp{\left( - \frac{\Delta F_c}{T} \right)}\ .
\label{10}
\end{equation}
Here $I_0$ is the preexponential factor and $\Delta F_c$ is the excess 
free energy of the critical cluster in the system. 
In Langer's theory the prefactor $I_0$ is a product of a dynamical and 
a statistical prefactor, $\kappa$ and $\Omega_0$, respectively
\begin{equation}
\displaystyle
I_0 = \frac{\kappa}{2 \pi} \Omega_0\ .
\label{11}
\end{equation}
It is interesting that under certain assumptions the prefactor 
derived in the classical theory may be obtained \cite{BrZa97a} 
identical to that of the modern theory.

To clarify the meaning of both the dynamical and the statistical 
prefactors, let us consider a classical system with $N$ degrees of 
freedom described by a set of $N$ 
collective coordinates $\eta,\ i = 1, \ldots , N$. The coarse-grained 
free energy functional $F\{\eta\}$ of the system has local minima 
$F\{\eta_i\}$ in the $\{\eta\}$-space, corresponding to the metastable 
and stable states, separated by the energy barrier. The point of 
minimal energy along the barrier is the so-called saddle point 
$\{\eta^S\}$. Note that this saddle-point configuration corresponds to 
the critical cluster of a condensing phase in the classical theory. In 
contrast, in field theory the critical cluster of a condensing phase
may not necessarily be a physical object but rather corresponds
to a certain saddle-point configuration in phase space. 
The phase transition occurs when the configuration $\{\eta_i\}$ moves 
from the vicinity of a metastable minimum to the vicinity of a stable 
one. When the potential barrier is overcome, it is most likely for 
the trajectory of the system to pass across a small area around the 
intermediate saddle point $\{\eta^S\}$. The rate of the decay of the 
metastable state is determined by the steady-state current across the 
saddle point from the metastable to the stable minimum of $F\{\eta\}$.

Performing the Taylor series expansion of $F\{\eta\}$ around the 
$\{\eta^S\}$ and keeping only quadratic terms yields
\begin{equation}
\displaystyle
F\{\eta\} - F\{\eta^S\} = \frac{1}{2} \sum_{k=1}^{N} \lambda_k 
(\eta_k - \eta_k^S)^2 \ ,
\label{12}
\end{equation}
where $\lambda_k$ are the eigenvalues of the matrix 
$M_{ij} = \partial ^2 F / \partial \eta_i \partial \eta_j$, 
evaluated at the saddle point. By definition, at $\{\eta^S\}$ the free 
energy density functional reaches its local maximum. Therefore, at 
least one of the eigenvalues $\lambda_k$ must be negative. Following 
Langer, we denote it as $\lambda_1$.
Then one may approximate the potential barrier between the 
metastable and stable well by the excess of the Helmholtz free energy,
$\Delta F$, corresponding to the formation of a spherical bubble of
size $R$. In the thin wall approximation, $\Delta F$ is the sum 
of the bulk and the surface energies,
\begin{equation}
\displaystyle
\Delta F(R) = - \frac{4\pi}{3}R^3\Delta p \, +\, 4\pi R^2 \sigma \ ,
\label{13}
\end{equation}
with $\Delta p$ being the difference in pressures inside and outside 
of the bubble. In the droplet model of Fisher \cite{Fish67}, the 
activation free energy includes also the so-called curvature term. It 
arises due to the small fluctuations in the shape of the bubble, which 
leave unchanged both, the volume and the surface area of the bubble,
\begin{equation}
\displaystyle
\Delta F^F(R) = -\frac{4}{3} \pi R^3 \Delta p + 4 \pi \sigma R^2
 + 3 \tau T \ln{\frac{R}{r_0}} \ .
\label{14}
\end{equation}
Here $\tau \approx 2.2$ is the Fisher critical exponent and 
$r_0$ is the radius of the smallest bubble in the system. 
Minimization of $\Delta F$ with respect to the radius $R$ yields 
the free energy of the critical bubble
\begin{equation}
\displaystyle \Delta F^F_c = \frac{4}{3} \pi R_c^2 \sigma
+ \tau T \left( 3 \ln{\frac{R_c}{r_0}} - 1 \right) \ .
\label{15}
\end{equation}
Here the critical radius $R_c$ should be evaluated by solving the
equation $\displaystyle \partial \Delta F^F / \partial R = 0$. 

It is convenient to introduce new variables \cite{BrZa95}: the 
similarity number,
$\displaystyle \lambda_Z = R_c \, \sqrt{ 4 \pi \sigma /T } $, 
and the reduced radius, $r = R/\!R_c$. In terms of these variables 
we have
\begin{eqnarray}
\displaystyle
\frac{ \Delta F^F_c}{T} &=& -\tau + \frac{\lambda_Z^2}{3} + 3 \tau 
\ln{\frac{R_c^F}{r_0} } \\
\label{16}
\frac{\Delta F^F}{T} &=& - \left( \tau + \frac{2}{3} \lambda_Z^2 
\right) r^3 + \lambda_Z^2 r^2 + 3 \tau \ln{\frac{R_c^F}{r_0} } \ .
\label{17}
\end{eqnarray}

In the harmonic approximation for the activation energy of a bubble 
near the critical radius, Eq.(\ref{17}) reads
\begin{eqnarray}
\displaystyle
\frac{\Delta F^F}{T} &=& \left( \frac{\Delta F^F}{T} \right)_{R=R_c} +
\frac{1}{2 T} \left( \frac{\partial^2 \Delta F^F}{\partial R^2}
\right)_{R=R_c} (R - R_c)^2 \\ 
\nonumber
 &=& \frac{ \Delta F^F_c}{T} - \frac{9 \tau + 2 \lambda_Z^2}{2}
 (r - 1)^2 \ ,
\label{18}
\end{eqnarray}
and we get finally for the only negative eigenvalue $\lambda_1$:
\begin{equation}
\displaystyle
\lambda_1 = - T \left( 9 \tau + 2 \lambda_Z^2 \right) \ .
\label{19}
\end{equation}
This expression will be used to determine the nucleation rate of the
process, in particular the statistical prefactor.

\subsection{Dynamical prefactor}
\label{subsec3a}

The dynamical prefactor, $\displaystyle \kappa = {\rm d}/{\rm d}t 
\left[ \ln (R - R_c) \right] $, which is related to the single 
negative eigenvalue $\lambda_1$, determines the growth rate of the 
critical bubble of size $R_c$ at the saddle point. To compute $\kappa$ 
one has to solve the hydrodynamic equations \cite{CsKa92a,TuLa80}
which describe the growth of a bubble of the hadronic phase due to the 
diffusion flux through the interface.

For a baryon-free plasma, where the thermal
conductivity is absent because of the absence of a rest frame
defined by the baryon net charge, the dynamical prefactor has been
calculated by Csernai and Kapusta \cite{CsKa92a} to
\begin{equation}
\displaystyle
\kappa_1 = \frac{4 \sigma \left( \zeta + \frac{4}{3} \eta \right)}
{(\Delta \omega)^2\, R_c^3}\ ,
\label{20}
\end{equation}
containing the bulk and the shear viscosities, $\zeta$ and $\eta$, 
and the difference $\Delta \omega$ between the enthalpy
densities of the plasma and hadronic phase, $\omega = e + p$.

Here it is implied that the energy flow, $\omega \vec{v} $,
where $\vec{v}$ is the velocity of the net particles, is provided
by viscous effects. Recently Ruggeri and Friedman \cite{RuFr96}
argued that the energy flow does not vanish even in the absence
of heat conduction. Since the change of the energy density $e$
in time is given in the low velocity limit by the conservation
equation,
\begin{equation}
\displaystyle
\frac{\partial e}{\partial t} = - \nabla \cdot \left( 
\omega \vec{v} \right)\ ,
\label{21}
\end{equation}
this means that the energy flow $\propto \omega \vec{v}$ is always 
present. Then the calculation of the dynamical prefactor for a 
system with zero thermal conductivity leads to the expression  
\begin{equation}
\displaystyle
\kappa_2 = \left( \frac{2 \sigma \omega_q}{R_c^3 (\Delta \omega)^2}
\right)^{1/2}
\label{22}
\end{equation} 
which agrees with the result of \cite{LaTu73} obtained for nonviscous
systems, i.e. the viscous effects cause only small perturbations to
Eq.~(\ref{22}). Note that the dynamical prefactor $\kappa_2$ might
violate the dynamical scaling laws of Kawasaki \cite{Kawa75} in the 
vicinity of the critical point (for details see 
\cite{TuLa80,LaTu73,Kawa75}) and, therefore, should be handled with
great care.

\begin{figure}[htp]
\centerline{\epsfysize=15cm \epsfbox{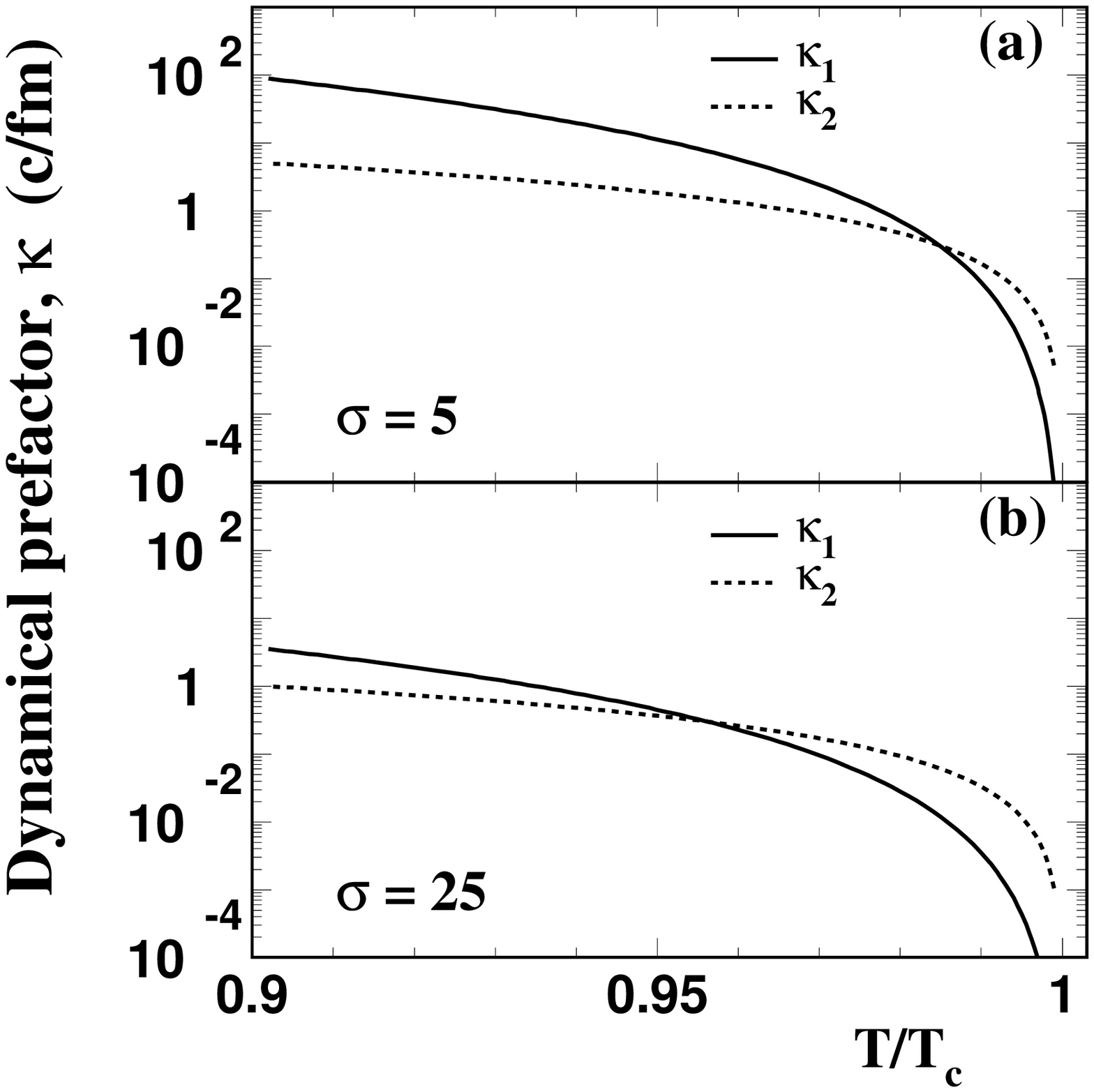}}
\caption{ 
Dynamical prefactors $\kappa_1$ (solid line) and $\kappa_2$ (dashed
line) calculated with $\sigma = 5\,$MeV/fm$^2$ {\bf (a)} and 
$\sigma = 25\,$MeV/fm$^2$ {\bf (b)} vs the temperature of the system.
}
\label{fig1}
\end{figure}

Figure~\ref{fig1} shows the evolution of the dynamical prefactors, 
$\kappa_1$ and $\kappa_2,$ below the critical temperature for 
different values of the surface tension. 
To calculate $\kappa_1$ we use the fact that the shear viscosity of
a two-flavored QGP, $ \eta = 1.29\, T^3 / [\alpha_S^2\, \ln( 1/
\alpha_S )]$ \cite{Heis94}, is much larger as compared with the bulk 
viscosity $\zeta$. One can see that Eq.~(\ref{22}) predicts
higher rates $\kappa$ for moderate or weak supercooling. It means,
in particular, that the minimal temperature, reached by the system 
during the cooling stage, should be higher if $\kappa_2$ is used in 
the calculations rather than $\kappa_1$. We will discuss the effect 
of the replacement of the dynamical prefactor on the course of the 
plasma hadronization in the model in Sec.~\ref{sec4}. 

Homogeneous nucleation theory permits us also to determine the 
macroscopic radial velocity of bubbles \cite{BrZa97a} via the
dynamical prefactor, $\kappa$, and the critical radius, $R_c$. 
Assuming that bubbles of hadronic phase grow due to the diffusion 
flux through their interface we have
\begin{equation}
\displaystyle
v(R) \equiv \frac{{\rm d} R}{{\rm d} t} = |\kappa|\,\left(
\frac{R_c}{R} \right)^2\, (R - R_c)\ .
\label{23}
\end{equation}
If $R = R_c$, the radial velocity drops to zero and the bubble is in
(metastable) equilibrium. If $R < R_c$, $v(R)$ is negative and, hence, 
the bubble collapses. If $R > R_c$, the radial velocity is positive 
and the bubble grows.

\subsection{Statistical prefactor}
\label{subsec3b}

The statistical prefactor, $\Omega_0$, is a measure of the volume of 
the saddle-point region in phase space available for nucleation.
Sometimes $\Omega_0$ is called also generalization of the Zeldovich
factor $Z$ \cite{Fren46}, although this is a crude simplification, 
since the relation between these two factors is actually more complex
\cite{BrZa97a}. The product of $\Omega_0$ and the exponential
$\displaystyle \exp \left( - \Delta F_c / T \right)$ gives the
probability of finding the system at the saddle-point$-$ rather than 
at the metastable configuration.

According to \cite{Lang67,Lang69,CaCo77,Affl81}, the statistical 
prefactor can be written as
\begin{equation}
\displaystyle
\Omega_0 = {\cal V}\, \left( \frac{2 \pi T}{|\lambda_1|} \right)
^{1/2}\, \left[ \frac{{\rm det}\,(M_0 / 2 \pi T)}{{\rm det}\,
(M^\prime /2 \pi T)} \right]^{1/2}\ ,
\label{24}
\end{equation}
where ${\cal V}$ is the available phase space volume at the saddle
point, the index "0" denotes the metastable state, and the prime 
indicates that the negative eigenvalue $\lambda_1$, as well as the 
zero eigenvalues of the matrix $M_{ij}$, is omitted.

The calculation of the fluctuation determinant in Eq.~(\ref{24}) is
usually extremely difficult and, moreover, a very important 
uncertainty exists in the determination of $\Omega_0$.
Indeed, in the harmonic approximation (\ref{12}) for the free energy
density functional $F\{\eta\}$ 
\begin{equation}
\displaystyle
\Omega_0 = {\cal V}\, \left( \frac{2 \pi T}{|\lambda_1|} \right)
^{1/2}\, \prod_{l=l_0+2}^N \left( \frac{2 \pi T}{\lambda_l^{(S)}}
\right)^{1/2} \, \prod_{l=1}^N \left( \frac{\lambda_l^{(0)}}{2 \pi T}
\right)^{1/2} \ . 
\label{25}
\end{equation}
Here $\lambda_l^{(S)}$ and $\lambda_l^{(0)}$ are eigenvalues of the 
mobility matrix $M$, evaluated at the saddle point and at the 
metastable point, respectively, and $l_0$ is the total number of 
symmetries of $\{F\}$ which are broken by the presence of the
saddle-point configuration. Since it is the translational symmetry
of the system that is broken due to the bubble creation, the three
translation invariance zero modes omitted in the fluctuation 
determinant give rise to the prefactor proportional to 
$\lambda_1^{-3/2}$ in the expression for the available phase space
volume ${\cal V}$:
\begin{equation}
\displaystyle
{\cal V}\, = V \left( \frac{8 \pi \sigma}{3 |\lambda_1|}
\right)^{3/2} \ .
\label{26}
\end{equation}
Here $V$ is the total volume of the system. In \cite{Lang69} the
fluctuation corrections in the products over $\lambda_l^{(S)}$ and
$\lambda_l^{(0)}$ from (\ref{25}) are absorbed into the free energy
of the metastable region and the saddle-point region, namely
\begin{eqnarray}
\displaystyle
\exp{\left( - F_0 / T \right)} &\equiv & 
\exp{\left( -F\{\eta^0\} / T \right)} \,
\prod_{l=1}^N \left( \frac{2 \pi T}{\lambda_l^{(0)}} \right)^{1/2} 
\ , \\
\label{27}
\exp{\left( - F_S / T \right)} &\equiv &
\exp{\left( -F\{\eta^S\} / T \right)}
 \prod_{l=l_0+2}^N \left( \frac{2 \pi T}{\lambda_l^{(S)}} 
 \right)^{1/2} \ .
\label{28}
\end{eqnarray}
Therefore, the activation energy of a critical cluster is simply
$\Delta F_c = F_S - F_0$ and
\begin{equation}
\displaystyle
\Omega_0 = V\, \frac{32 \pi^2 T^{1/2}}{|\lambda_1|^2}\,
\left( \frac{\sigma}{3} \right)^{3/2} \ .
\label{29}
\end{equation}
In Ref.~\cite{LaTu73} it was mentioned that there are four more
terms in the product over $\lambda_l^{(0)}$ than in that over 
$\lambda_l^{(S)}$. Therefore, the free energy difference can not 
be precisely a logarithm of these products. The final expression for 
the statistical prefactor, which accounts for the four unpaired
$\lambda_l^{(0)}$'s modes, reads
\begin{equation}
\displaystyle
\Omega_0 = \frac {2}{3 \sqrt{3}} \, \frac{V}{\xi^3} \, 
\left( \frac{\sigma \xi^2}{T} \right)^{3/2} \,
\left( \frac{R_c}{\xi} \right)^{4} \ ,
\label{30}
\end{equation}
where $\xi$ is the correlation length.

\begin{figure}[htp]
\centerline{\epsfysize=15cm \epsfbox{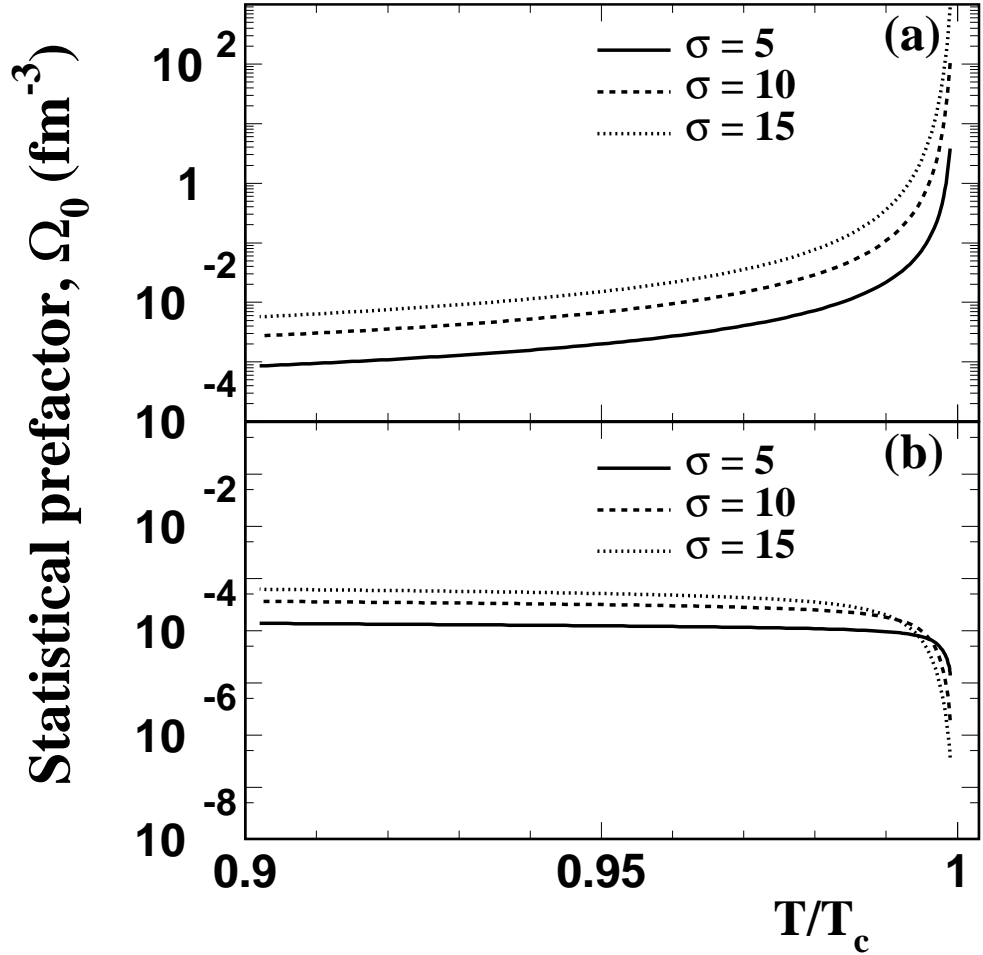}}
\caption{ 
{\bf (a)}: Statistical prefactor $\Omega_0$ given by 
Eq.(\protect \ref{30}) vs temperature with $\sigma = 5$
(solid line), 10 (dashed line), and 15 (dotted line) MeV/fm$^2$.
{\bf (b)}: The same as {\bf (a)} but for 
Eq.(\protect \ref{29}).
}
\label{fig2}
\end{figure}

The statistical prefactors corresponding to the expressions 
(\ref{29}) and (\ref{30}) are shown in Fig.~\ref{fig2}. In
contrast to the curves in the lower panel, the curves in the upper 
panel diverge at $T = T_c$. This is an unphysical result, which is 
due to the fact that the nucleation process is turned off as $T 
\rightarrow T_c$.It is worth noting, however, that the rise of 
$\Omega_0$ is counterbalanced by the factor $\exp{(-\Delta F_c/T)}$ 
that drops to zero at the critical temperature. Therefore, the total 
nucleation rate will increase with dropping temperature for both 
prefactors. Also, the correlation length, $\xi$, is not a constant, 
but rather scales with the temperature in the proximity of the 
critical point, where the PT turns to second order, as $\displaystyle 
\xi(T) = \xi(0) \left( 1 - T_{\rm crit}/T \right)^{-\nu} $, with the 
critical exponent $\nu = 0.63$ \cite{LaTu73}. Since such a critical 
point does not exist in the plasma to hadron gas phase transition 
scheme presented here, Eq.~(\ref{29}) is used to calculate the 
statistical prefactor $\Omega_0$.

\section{RELAXATION OF THE METASTABLE QGP}
\label{sec4}

Like in any phenomenological model, the scenario of a homogeneously 
nucleating QGP is based on a set of model parameters. Here, the role 
of each parameter for the dynamics of the PT is studied in order to 
reveal those most relevant parameters to which the model results are
most sensitive.

\begin{figure}[htp]
\centerline{\epsfysize=15cm \epsfbox{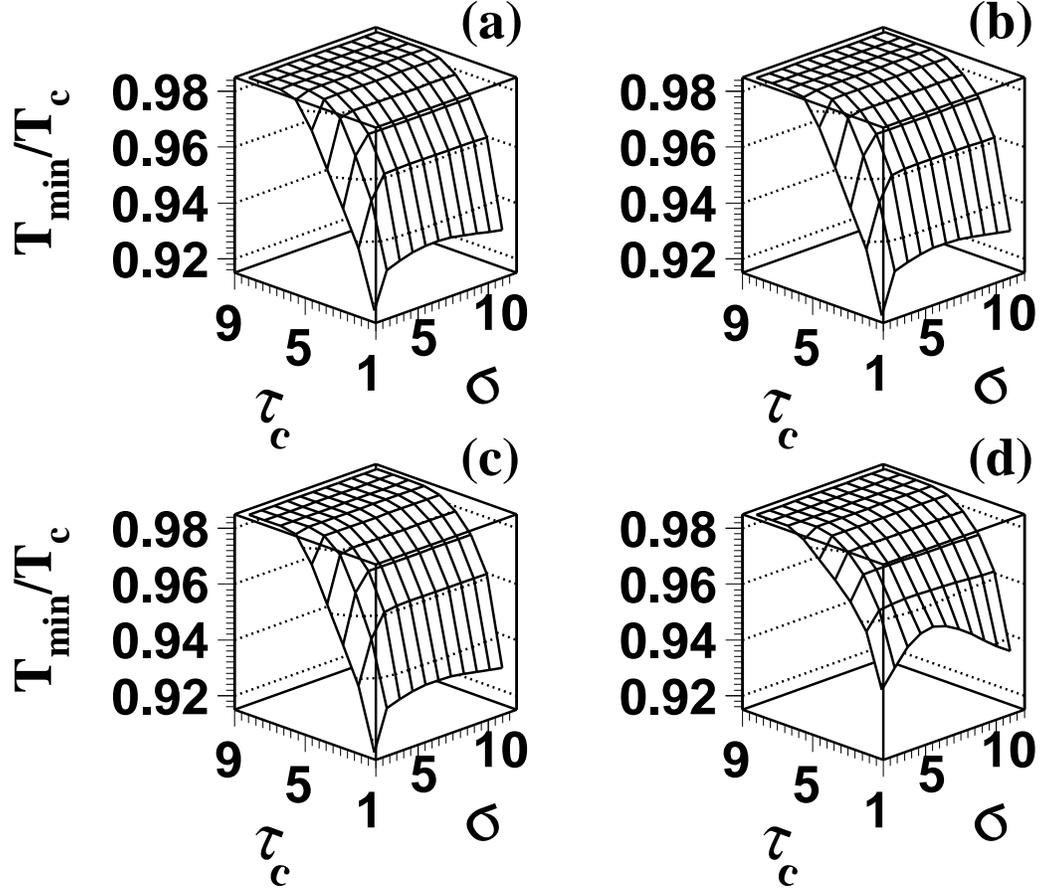}}
\caption{ 
Supercooling of the system as function of critical time $\tau_c$ and 
surface tension $\sigma$.
{\bf (a)}: bubbles grow together with the total volume (scaling 
regime), entropy of the bubble surface is neglected.
{\bf (b)}: bubbles grow independently on the total volume (non-scaling
regime), entropy of the bubble surface is neglected.
{\bf (c)}: non-scaling regime, surface entropy is taken into account.
{\bf (d)}: the same as {\bf (c)}, but prefactor $\kappa_1$ is 
substituted by $\kappa_2$.
}
\label{fig3}
\end{figure}

The theory of homogeneous nucleation is valid for systems which 
are not too far from equilibrium. In particular, the supercooling 
in the system should not be too strong $-$ otherwise the nucleation
theory fails! The minimum temperature reached by the expanding and
cooling plasma is shown in Fig.~\ref{fig3} as a function of the 
surface tension, $\sigma$, and the critical time, $\tau_c$. The most 
reliable value of the surface tension lies within the range $0.015 
\leq \sigma/T_c \leq 0.1$ \cite{Iwa94,Kar96}, i.e. $2 \leq \sigma 
\leq 12.5\,$MeV/fm$^2$ for the given $T_c$. Note that even in the
case of an extremely fast expansion and very low values of $\sigma$,
the supercooling of the system does not exceed 8\%. This result does 
practically not change when bubbles grow independently in the 
non-scaling regime on the growing total volume [Fig.~\ref{fig3}(b)].
Even when the surface tension and pressure of the bubble surface is 
included in the expressions for total energy density and pressure,
Eqs.~(\ref{4})$-$(\ref{5}), [Fig.~\ref{fig3}(c)], the results stay
put. 

If the dynamical prefactor given by Eq.~(\ref{20}) is replaced by that 
of Eq.~(\ref{22}), the amount of the QGP volume converted into hadrons 
at the earlier stages of the supercooling is increased. Thus, the 
supercooling of the system in the latter case is about 2\% weaker 
[Fig.~\ref{fig3}(d)] than that shown in Fig.~\ref{fig3}(a)$-$(c).

In the analysis of the QGP equilibration times within the parton 
cascade model (PCM) by Geiger \cite{Geig92}, which has been done for 
the RHIC and the LHC energies, yields the earliest 
equilibration time for the plasma as $\tau_{\rm init} = 3/8\,$fm/$c$ 
with $T_{\rm init} \approx 2 T_c$. Then the critical temperature
will be reached at $\tau_c = 3\,$fm/$c$, which corresponds to a
supercooling of about 5$-$6\% in our figures. Therefore, one may 
conclude that the application of the homogeneous nucleation theory to 
the hadronization process of even a relativistic expanding QGP seems 
quite reasonable. 

\begin{figure}[htp]
\centerline{\epsfysize=15cm \epsfbox{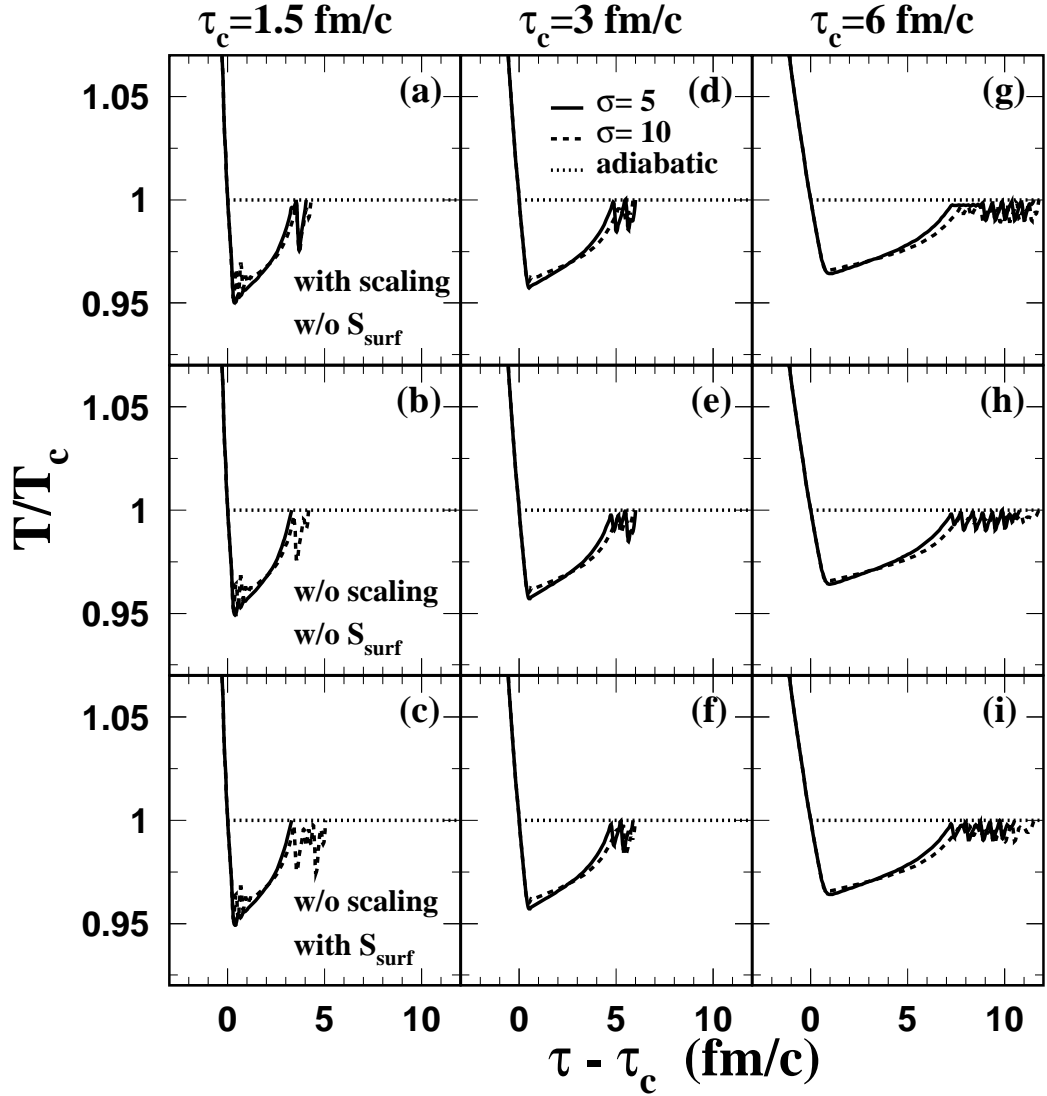}}
\caption{ 
The temperature evolution with time for the expansion 
scenarios with $\tau_c = 1.5\,$fm/$c$ [{\bf (a)-(c)}], 3$\,$fm/$c$
[{\bf (d)-(f)}], and 6$\,$fm/$c$ [{\bf (g)-(i)}]. The sequence of
conditions within each subgroup of panels corresponds to that of
Fig.\protect \ref{fig3} {\bf (a)-(c)}. The dotted curves are the 
idealized adiabatic scenario of the PT, the others show
the result of solving of rate equations with $\sigma = 5$ (solid 
lines) and 10 (dashed lines) MeV/fm$^2$.
}
\label{fig4}
\end{figure}

The temperature is plotted as a function of the proper time $\tau$
in Fig.~\ref{fig4} for $\tau_c = 1.5,\ 3\ {\rm and}\ 6\,$fm/$c$. The 
upper panels in this figure corresponds to conditions (a) of 
Fig.~\ref{fig3}, the middle row corresponds to conditions (b), and the 
lower row corresponds to conditions (c) of the same figure.
Hadronization causes the release of the latent heat, and the
system reheats to temperatures close to $T_c$. Then the nucleation 
and growth of hadronic bubbles comes to a halt. 

The continuing increase of the total volume leads again to a decrease 
of the temperature, and the phase transition continues immediately. 
These oscillations of the temperature in the vicinity of $T_c$ are 
well observable in Fig.~\ref{fig4}. The mixed system is quite unstable 
at this stage, since any negligibly small rise of the temperature 
forces the system to reach the critical point, where, in turn, it may 
break-up into fragments. Then the theory of spinodal decomposition 
might be relevant to describe the hadronization of the rest of the 
QGP. The time $\Delta \tau$ needed for the system (for $\sigma = 
5-10\,$MeV/fm$^2$) to reach the zone of the oscillations scales with 
$\tau_c$ as
\begin{equation}
\displaystyle
\Delta \tau = 2.9\, \tau_c^{1/2}\ ,
\label{30a}
\end{equation}
within the interval $1.5 \leq \tau_c \leq 9\,$fm/$c$.

\begin{figure}[htp]
\centerline{\epsfysize=15cm \epsfbox{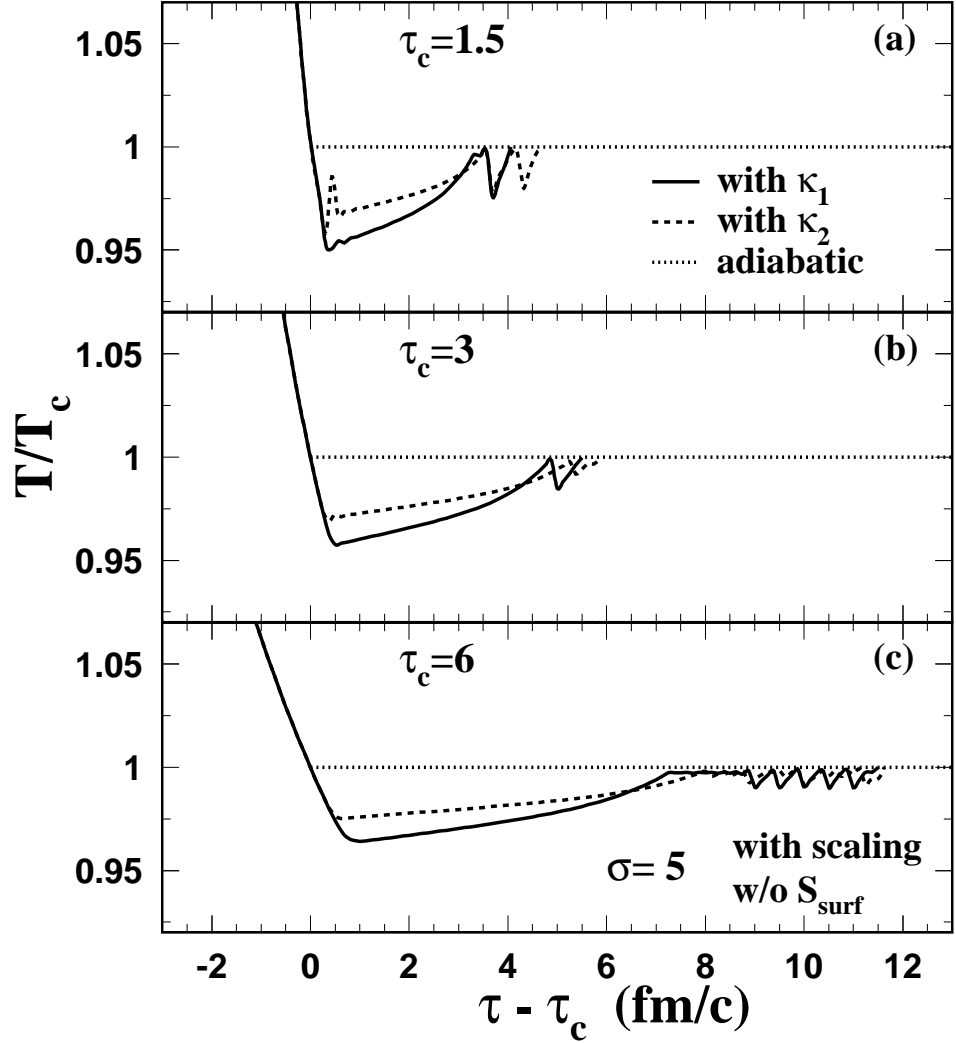}}
\caption{ 
{\bf (a) - (c)}: the same as Fig.\protect \ref{fig4} {\bf (a),(d),
(g)} with $\sigma = 5\,$MeV/fm$^2$ but for the solutions of the rate 
equations with $\kappa_1$ (solid lines) and with $\kappa_2$ (dashed 
lines).
}
\label{fig5}
\end{figure}

Changing the dynamical prefactor from $\kappa_1$ to $\kappa_2$
(Fig.\ref{fig5}) leads to negligibly small shifts in the time needed
to reach the vicinity of the critical temperature. This is due to the
fact that the late time evolution of the system is governed by the
Lifshitz-Slyozov dynamics, which is generally much more important
\cite{LiSl61} for the course of a first order PT than the initial 
size distribution of clusters given by the nucleation theory. 

Therefore, the 
model does not appear to be very sensitive to a non-scaling growth of 
the nucleated bubbles, to the incorporation of the surface entropy in 
the rate equations, and to the numerical values of the dynamical 
prefactors obtained within the range of model parameters applied. The 
most important parameters are the value of the surface tension
$\sigma$ and the time to reach of the transition temperature $\tau_c$, 
which is determined by the initial conditions and by the expansion 
dynamics of the system. The effects of varying just these two factors 
for the relaxation process of the metastable QGP are studied below.

\begin{figure}[htp]
\centerline{\epsfysize=15cm \epsfbox{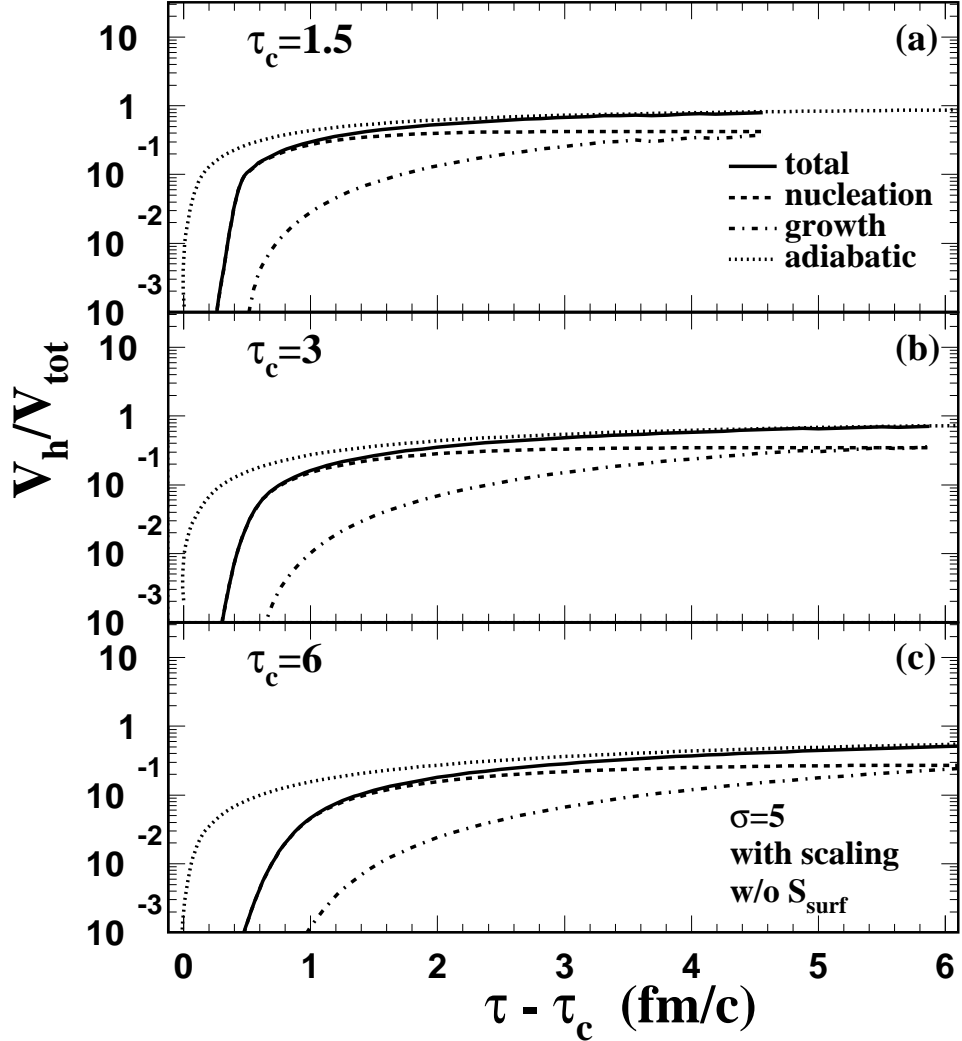}}
\caption{ 
The part of the QGP volume converted into hadrons for 
the idealized adiabatic PT (dotted lines)
and for homogeneous nucleation scenario with 
$\sigma = 5\,$MeV/fm$^2$.
Solid lines correspond to the total volume fractions of hadronic
matter, dashed lines denote the increase of hadronic volume due to 
the nucleation of new bubbles, and dash-dotted lines indicate the
enlargement of the hadronic bubbles due to diffusion.
}
\label{fig6}
\end{figure}

From Fig.~\ref{fig6} one may conclude that, at the very beginning of
the phase transition, the process of bubble nucleation is the main
mechanism of plasma conversion. The creation of the new phase reaches
its saturation value soon after 0.5$-$1.5$\,$fm/$c$, and the growth 
of the total hadronic fraction of the total volume proceeds due to the
diffusion growth of already nucleated bubbles. Both the nucleation and
the diffusion process contribute (almost equally) to the hadronization 
of the QGP.

\begin{figure}[htp]
\centerline{\epsfysize=15cm \epsfbox{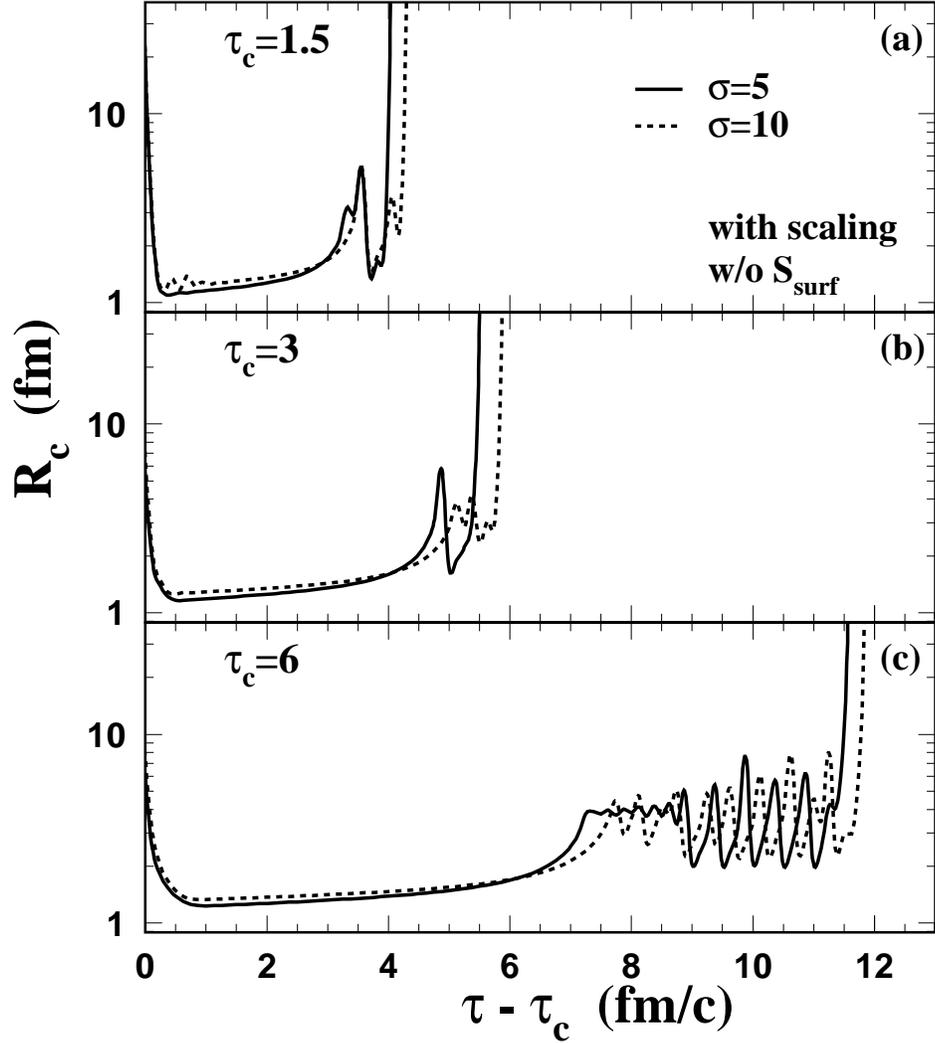}}
\caption{ 
Change of the critical radius as a function of time $\tau - \tau_c$ 
for the expanding system with $\tau_c = 1.5$ {\bf (a)}, 3 {\bf (b)},
and 6 {\bf (c)} fm/$c$, with $\sigma = 5$ (solid lines) and 10 
(dashed lines) MeV/fm$^2$.
}
\label{fig7}
\end{figure}

\begin{figure}[htp]
\centerline{\epsfysize=15cm \epsfbox{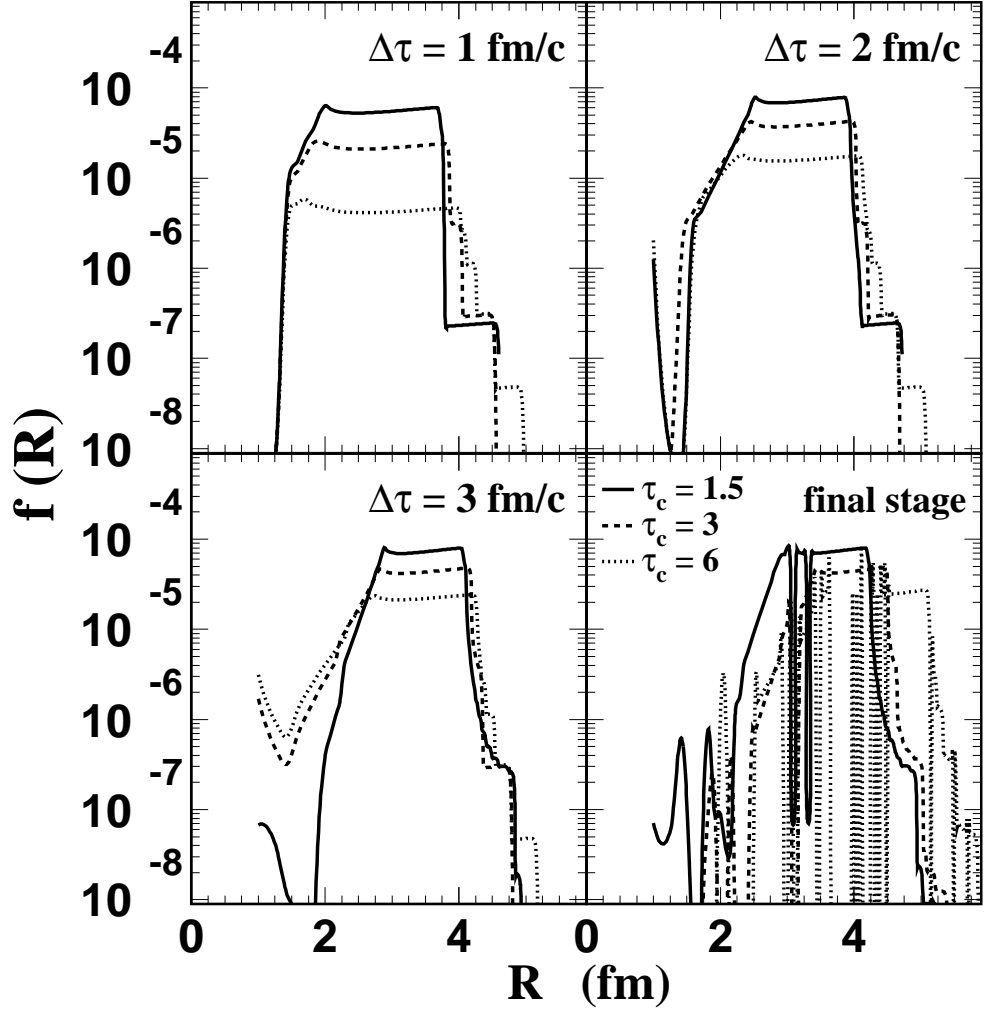}}
\caption{ 
Size distribution of hadronic bubbles at $\Delta \tau =
1, 2, {\rm and}\ 3\,$fm/$c$ after the beginning of the nucleation,
and at the freeze-out for $\tau = 1.5,\ \Delta \tau = 4\,$fm/$c$ 
(solid curves), $\tau = 3,\ \Delta \tau = 6\,$fm/$c$ (dashed curves),
and $\tau = 6,\ \Delta \tau = 11.5\,$fm/$c$ (dotted curves).
}
\label{fig8}
\end{figure}

The critical radius in the system initially drops to about 1$\,$fm 
(Fig.\ref{fig7}), then, due to reheating, it rises up to $R_c \approx
2-4\,$fm. Then the oscillations begin. At $T = T_c$, the critical
radius is, of course, infinite.  The weak changes of the temperature 
near the critical point cause significant oscillations in the value of 
the critical radius which, in turn, are responsible for the 
irregularities in the size distribution of bubbles at the final stage 
of the PT (see Fig.~\ref{fig8}, right lower panel).
The radial oscillations of the intermediate-sized hadronic bubbles can
result in the pulsed emission of matter and radiation owing to the
strong acceleration and deceleration of matter in the vicinity of the 
bubble surface. These oscillations are analogous to the pulsations of 
a hot quark blob, discussed in \cite{SGMG80}. Note also, that the 
bubble pulsations, occurring in medium, generate the sonic waves in 
the expanding plasma, but the discussion of this topic lies out of 
scope of the present paper.

When the nucleation just begins, the size distribution of hadronic
bubbles shown in Fig.~\ref{fig8} has a characteristic plateau-like
profile. At $\Delta \tau = 1\,$fm/$c$, just after entering the
metastable region, the critical radius drops to its minimal value of
about 1$\,$fm. Then practically all bubbles are growing. Reheating
of the system leads to the rise of the value of the critical radius.
As a result, a noticeable fraction of shrinking bubbles appears. 
The central plateau becomes narrower, also because of the increase of 
bubble density per unit of radial interval. At the end of the 
homogeneous nucleation stage, as mentioned above, the dips and peaks 
in the bubble size distribution arise due to the temperature 
oscillations. Note that the bubble size distribution established in 
the first order PT deviates clearly from the power-law distribution 
$f(A) \propto A^{-\tau}$ ($A$ being the size of a cluster), which is 
typical for the second order PT.

\begin{figure}[htp]
\centerline{\epsfysize=15cm \epsfbox{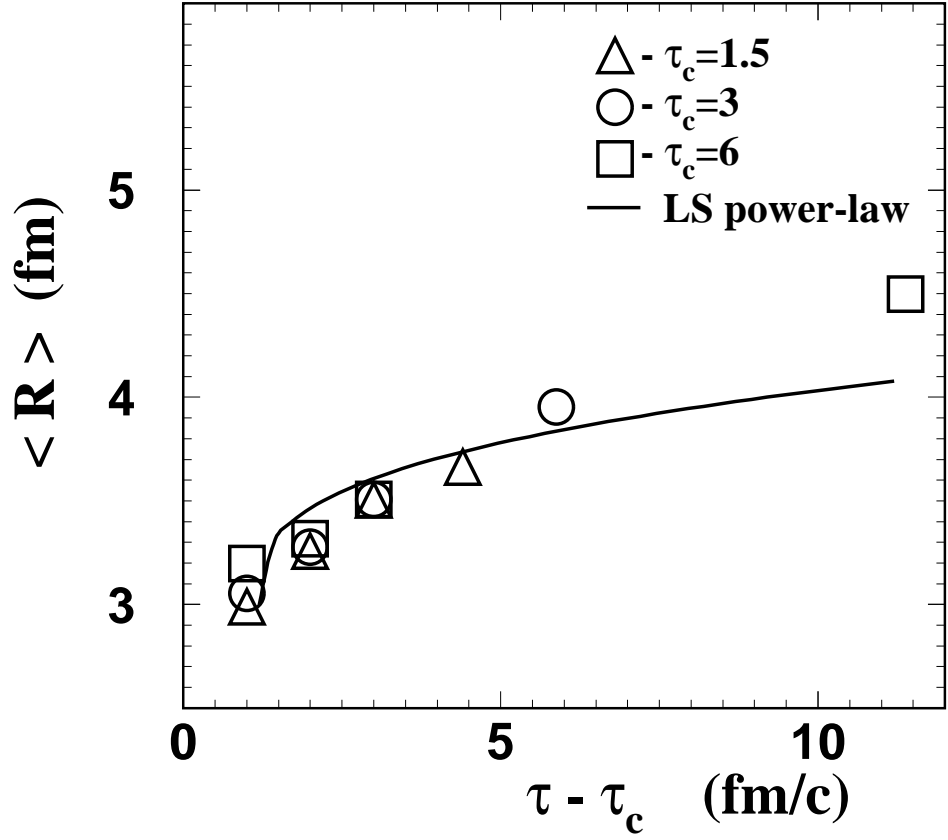}}
\caption{ 
Average radius of hadronic bubbles as a function of $\tau - \tau_c$
for the longitudinal expansion with $\sigma = 5\,$MeV/fm$^2$ and with
$\tau_c = 1.5$ (triangles), 3 (circles), and 6 (squares) fm/c. Solid
curve is the fit to Lifshitz-Slyozov power-law $<R> \propto t^{1/3}$.
}
\label{fig9}
\end{figure}

The scaling of the change of the average radius of the hadronic 
bubbles with time (Fig.~\ref{fig9}) demonstrates that the average 
radius depends mainly on the duration of the phase transition, but 
not on the expansion scenario. Since the nucleation of new bubbles is 
turned off after $\tau_0 \approx 1.2\,$fm/$c$, we fit the distribution 
to the power-law
\begin{equation}
\displaystyle
<R(\tau)> - <R(\tau_0)>\ =\ {\rm const} \times (\tau - \tau_0)^{1/3}
\label{31}
\end{equation}
with $R(\tau_0) = 3\,$fm, which is the Lifshitz-Slyozov (LS) 
$t^{1/3} -$law \cite{LiSl61} of the coalescence process.
We see that at the coalescence stage of the plasma conversion the 
agreement with the LS law is good. The deviations from this law 
at the late stage of the hadronization of QGP are also caused by
the temperature fluctuations.
 
\begin{figure}[htp]
\centerline{\epsfysize=15cm \epsfbox{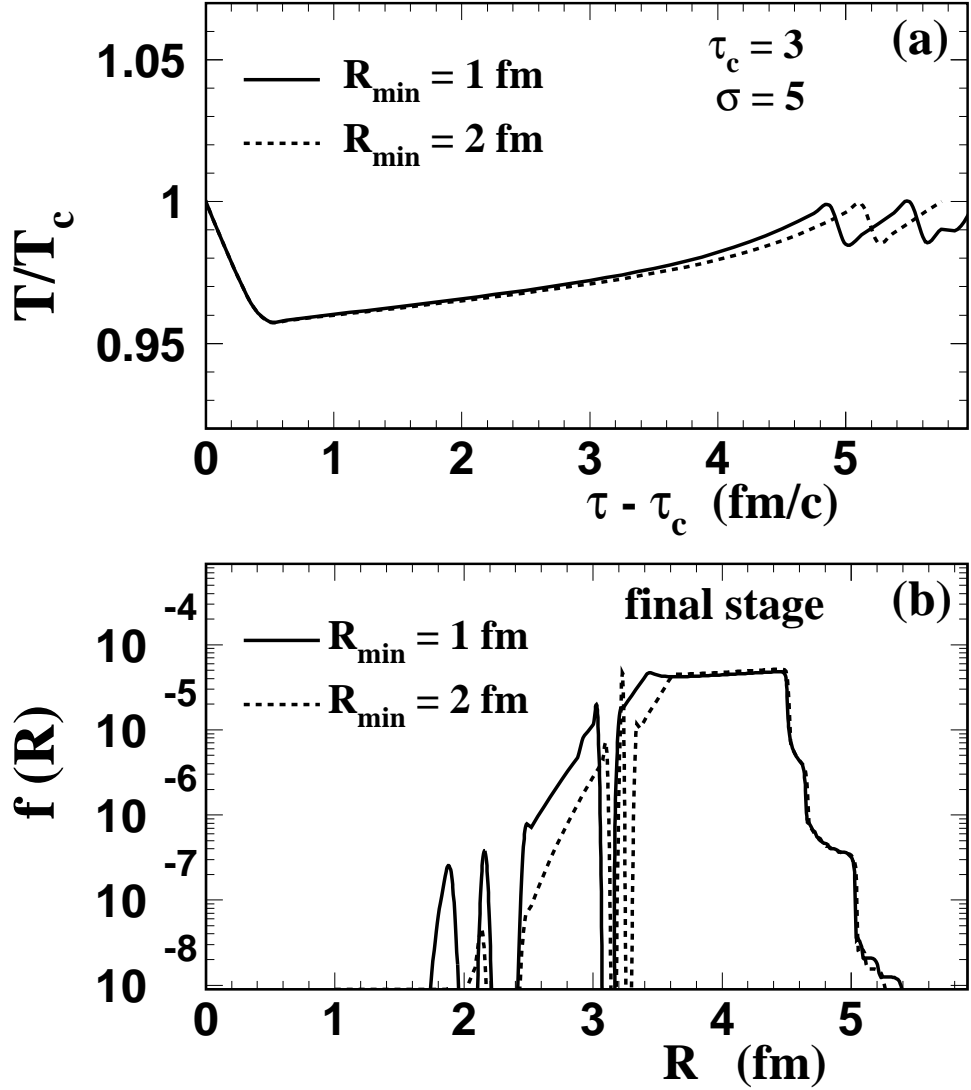}}
\caption{ 
{\bf (a)}: Temperature as a function of time for the expansion with 
$\sigma = 5\,$MeV/fm$^2$ and $\tau_c = 3\,$fm/c. Curves show the
result of solving of rate equations with minimum radius of hadronic 
bubbles $R_{min} = 1$ (solid line) and 2 (dashed line) fm.
}
\label{fig10}
\end{figure}

Finally, the dependence of the results on the minimum radius of the 
nucleated bubbles is studied. The upper panel of Fig.~\ref{fig10} 
depicts the temperature curves calculated with $r_0 = 1$ and $2\,$fm, 
and the lower panel compares the size distributions of the bubbles at 
the end of the PT. Again one can see that the effect of the cut-off 
of small bubbles is negligible because of the rather broad initial 
distribution of bubbles in size space.

The results of this Section may be summarized by concluding that the 
most important parameters of the model are the initial conditions, 
time, $\tau_{init}$, and temperature, $T_{init}$, of the QGP 
thermalization, as well as the scenario of further plasma expansion, 
and the value of the surface tension $\sigma$. All other factors cause 
only small deviations from the solutions obtained with the fixed set 
of the aforementioned parameters. Note that effects like the 
final transients near the critical point are sensitively dependent
on the fluctuations in the system, and these might (or might not)
wash out these transients. Real systems produced in experiments with
heavy ions are not infinite. Therefore, to complete our analysis, we
have to clarify the role of the finite size effects.

\section{ROLE OF THE FINITE SIZE EFFECTS}
\label{sec5}

The biggest problem which has to be overcome before a careful 
investigation of the finite size effects can be tackled is the 
problem of the first order phase transition itself. Then, a 
consistent treatment of the finiteness of the system and, especially, 
of the surface induced phenomena is still an open question. Attempts 
to estimate the role of the finite size effects on the phase 
transitions in nuclear matter have been made in 
\cite{GKM84,CsNe94,SSG98} on a basis of a purely thermodynamical
picture of fluctuations. Can this approach be modified and applied 
for our kinetic analysis of the plasma hadronization? To answer the 
question note that the situation we face here may be subdivided into 
two cases. 

\begin{figure}[htp]
\centerline{\epsfysize=15cm \epsfbox{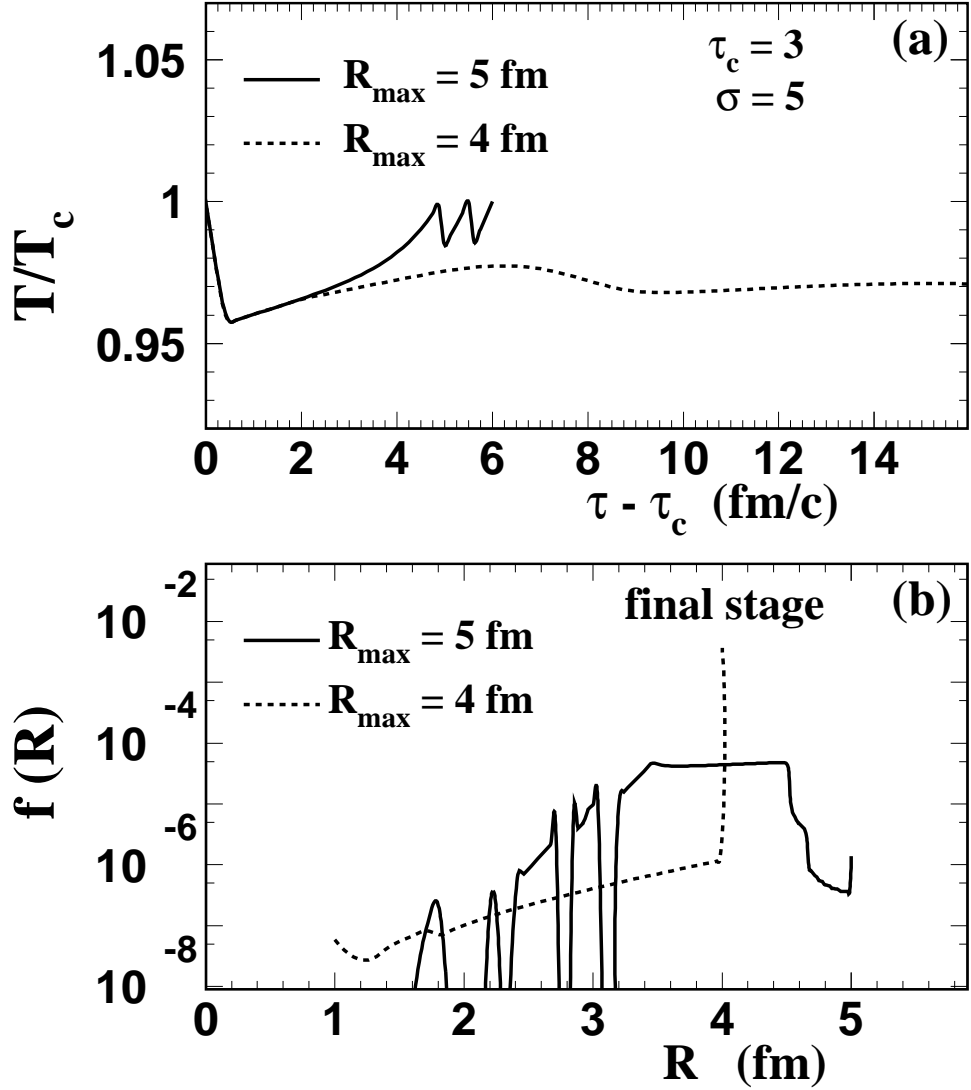}}
\caption{ 
{\bf Upper :} The same as Fig.\protect\ref{fig10} but for maximum
radius of hadronic bubbles $R_{max} = 5$ (solid line) and 4 (dashed
line) fm.
{\bf Lower :} Size distributions of hadronic bubbles at the freeze-out
for the same scenario as in upper panel 
with $R_{max} = 5$ (solid line) and 4 (dashed line) fm. 
}
\label{fig11}
\end{figure}

First, let 
us imagine that a droplet of plasma is immersed into a hot gas of 
hadrons which acts as a heat bath, i.e. the temperature fluctuations
in the droplet are suppressed. The compound system expands 
longitudinally, and the final size effects come into play via the
finiteness of the transverse direction. The results obtained in 
Sec.~\ref{sec4} are valid for central collisions of gold or lead
ions ($R \geq 7\,$fm) at relativistic energies. If then, by chance, 
the transverse radius of the expanding cylinder will be smaller than 
7$\,$fm, the cut-off of large bubbles should reduce the volume 
fraction occupied by hadrons and affect the course of the phase 
transition. Figure~\ref{fig11} presents the evolution of both the
temperature and the bubble size distribution as calculated with 
maximum radius 5$\,$fm and 4$\,$fm, respectively. The behavior of the 
system differs drastically between these rather close values of $R$:
as the central plateau in the size distribution lies within the range 
of $3 \leq r \leq 4.5\,$fm at $t \approx 10\,$fm/$c$, the cut-off of 
bubbles with $R > 5\,$fm does not cause noticeable deviations from 
the scenario discussed in Sec.~\ref{sec4}. However, for 
$R \leq 4\,$fm the maximum temperature reached by the system during 
the reheating is reduced and the conversion of the QGP into hadrons
is significantly delayed. As a result, the bubble size distribution 
has a pronounced peak at $R = 4\,$fm (Fig.~\ref{fig11}, lower panel). 

The values of the cut-off radius, at which the slowly varying 
temperature of the long-lived object does not exceed the $0.98\,T_c$
upper limit, are listed in Table~\ref{tab1} for various critical 
temperatures, surface tensions and expansion rates. It is easy to see 
that all values are within the range of the central plateau in the
bubble size distribution. Thus the most probable size of the emitting
sources is not affected by the finiteness of the system.
 
For radii smaller than $2-3\,$fm, the surface induced effects 
must be taken into account. Undoubtedly, the theoretical treatment of 
the interface between plasma and hadronic matter is oversimplified as 
compared with realistic system. Therefore we are not able to make any 
quantitative predictions, based on the homogeneous nucleation theory,
for the hadronization of such plasma filament. 

\begin{figure}[htp]
\centerline{\epsfysize=15cm \epsfbox{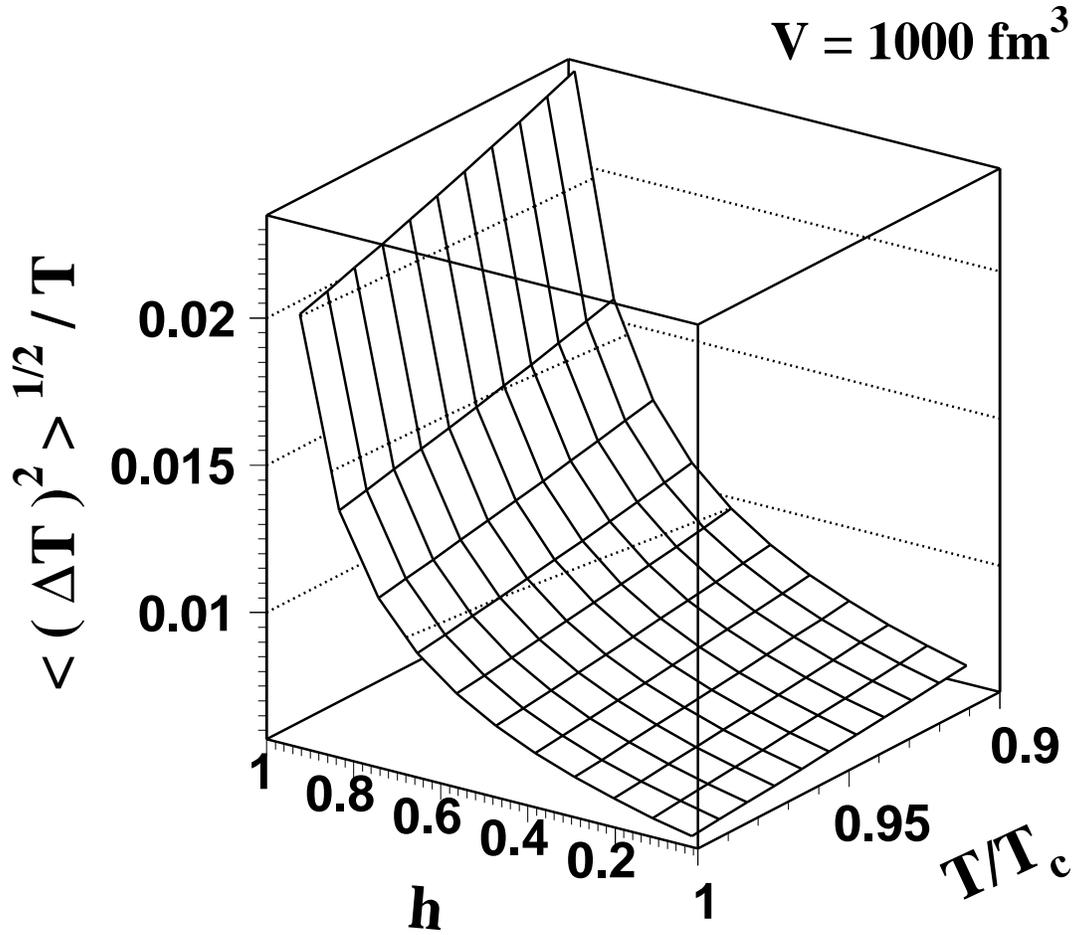}}
\caption{ 
Temperature fluctuations $\displaystyle (<(\Delta T)^2>)^{1/2}/T$ as
function of reduced temperature $T/T_c$ and hadronic fraction $h$ of 
the total volume of the system. 
}
\label{fig12}
\end{figure}

Second and perhaps more realistic is the case of the formation of a 
plasma pattern as an isolated system, i.e. without any contact with 
a heat reservoir. Although the energy of the expanding system is 
conserved, fluctuations of the temperature \cite{LaLi80} of the 
order of 
\begin{equation}
\displaystyle
<(\Delta T)^2> = \frac{T^2}{C_V} 
\label{32}
\end{equation}
should occur. Here $\displaystyle C_V = \left( \partial E/\partial T 
\right)_V$ is the heat capacity at constant volume. For the 
two-component system consisting of QGP and hadronic matter, the 
temperature $T$ is smeared around its mean value with the width 
\begin{equation}
\displaystyle
(<(\Delta T)^2>)^{1/2} = \frac{1}{2\, ( T\, V )^{1/2}}
\sqrt{\frac{1}{h\, a_h + (1-h)\, a_q}} \ .
\label{33}
\end{equation}
The smaller the volume $V$ of the system, the larger the temperature 
fluctuations, and vice versa. Moreover, from Fig.~\ref{fig12} one may
conclude that these fluctuations (at any given $T$) are larger in the 
hadronic system rather than that in the pure QGP phase. This is due to 
the smaller specific heat of the hadronic phase. We have seen in 
Sec.~\ref{sec4} that $-$ at temperatures of about 0.98$\, T_c$ (or 
lower) $-$ the critical radius and nucleation rate depend only very 
weakly on temperature variations within the range $ \Delta T / T_c 
\approx 2\,\%$. 

The situation changes completely when the temperature approaches the 
critical one. Since the hadronic matter here occupies already about 
$70\,\%$ of the total volume, fluctuations of the temperature should 
be as large as $1\,\%$ of $T_c$, i.e. 1.7$\,$MeV for $V = 
10^3\,$fm$^3$. For smaller volumes, say $10\,$fm$^3$, the fluctuations 
rise to $10\,\%$, while for a volume as large as 
$16 \cdot 10^3\,$fm$^3$ the width of the temperature smearing drops 
to $0.25\,\%$ of $T_c$, still almost fully covering the range of the 
temperature fluctuations. Thus the expanding mixed system should 
break-up into fragments even earlier than in the scenario with the 
presence of heat bath. The conclusions drawn in Sec.~\ref{sec4} remain 
true except for the appearance of irregularities in the size 
distribution of small and intermediate-size hadronic bubbles.

\section{DISCUSSION AND CONCLUSIONS}
\label{sec6}

It has been shown that the theory of homogeneous nucleation is 
applicable to describe the hadronization of a relativistically 
expanding quark-gluon plasma produced in heavy-ion collisions. We 
have proposed a coalescence-type model to follow further the evolution 
of hadronic bubbles produced in a metastable QGP. The change of the 
average radius of the bubbles with time is shown to be consistent with 
the LS power law $<R> \propto t^{1/3}$. 

Various sets of model dependent parameters are used to study the role
of each of them on the course of the plasma$-$hadrons PT. With rather 
good accuracy the number of these parameters may be reduced to the 
few main ones, namely initial conditions of plasma thermalization, 
expansion scenario, and the value of the surface tension of the 
interface
between plasma and hadronic matter. The supercooling of expanding
QGP is found to be relatively moderate, $5-6\,\%$ only. Then the 
system reheats up to the critical temperature, where the temperature
oscillations may occur. At $\Delta \tau = 5-10\,$fm/$c$ after the
beginning of nucleation process (this time depends strongly on the
initial conditions and model of expansion) the system hits the 
critical point. Since at that time already about $70-80\,\%$ of the
total volume is occupied by hadronic matter and the system is dilute
compare to initial state, the rest of the QGP may not be sufficient 
enough to "glue" the hadronic bubbles in a compound state. The 
expanding system simply breaks up into fragments: hadronic clusters 
and small droplets of plasma. Due to the finite size effects the
hadronization of QGP may be delayed and the long-lived objects
containing plasma and hadronic bubbles are produced. But the 
temperature fluctuations cause the broadening of the critical 
temperature region and earlier disintegration of the system.
One has to appreciate, however, that using the given EOS it is 
impossible to get the conversion of total amount of QGP faster than 
that of the idealized adiabatic transition. It means that the small
sources like the QGP droplets will burst, emitting hadrons, from time
to time up to about $40-50\,$fm/$c$, while the bulk amount of plasma 
is converted into hadrons within first $10\,$fm/$c$'s.

The entropy increase during the PT stage in the proposed
scheme is small, $2-7\,\%$, and there should be no significant 
changes in the integrated particle yields or pion to baryon ratios 
which can be detected experimentally. On the other hand, the formation
of large clusters and their further disassembly will lead to the 
significant multiplicity fluctuations in rapidity spectra of secondary
particles which should also have low transverse momenta. To search for
these clusters experimentally one can examine the rapidity and 
azimuthal distribution in small $p_T$-intervals, looking for islands
in sea of empty bins \cite{HLP94}. 

The nucleation process enforces the softening of EOS and diminishes
the transverse flow. Then, the cluster size distribution in the model
discussed departs from the simple power-law falloff, which is expected
for the second order PT. The most probable radius of emitting sources,
which is fully determined by the evolution of the value of critical 
radius with the temperature and by kinetics of the PT, varies from 
3$\,$fm to 4.5$\,$fm. This signal may be checked by the analysis of 
data on Hanbury-Brown-Twiss (HBT) correlations. However, the rest of 
the plasma dispersed between the hadronic bubbles is hadronizing also, 
giving rise to a substantial yield of small bubbles. Therefore, the 
presence of a plateau in the range $R = 3 - 4.5\,$fm in the size 
distribution of hadronic clusters can be considered as a signal of the 
first order phase transition. 

\section*{Acknowledgements} 

We thank L.P.~Csernai and M.I.~Gorenstein for the helpful discussions 
and valuable comments.
L.B. and E.Z. are indebted to the Institute for Theoretical Physics at 
the University of Frankfurt for hospitality.
This work was supported by the Graduiertenkolleg f{\"u}r Theoretische 
und Experimentelle Schwerionenphysik, Frankfurt$-$Giessen, the 
Bundesministerium f{\"u}r Bildung und Forschung, the Gesellschaft 
f{\"u}r Schwerionenforschung, Darmstadt, Deutsche 
Forschungsgemeinschaft, and the Alexander von Humboldt$-$Stiftung, 
Bonn.

\newpage
\mediumtext

\begin{table}

\caption{ Maximum transverse radius of longitudinally expanding
QGP at which the long-lived metastable state appears. The maximum
temperature reached by the system during the plasma conversion is 
less than $0.98\,T_c$.
}

\begin{tabular}{cccc}
Surface tension, &$ \tau_c = 1.5\,$fm/$c$ &$ \tau_c = 3\,$fm/$c$
&$ \tau_c = 6\,$fm/$c$  \\
$\sigma$ (MeV/fm$^2$)&$ R_{max}$ (fm) &$ R_{max}$ (fm) 
&$ R_{max}$ (fm)  \\
\tableline \tableline
 $ T_c\ =\ 150\,$MeV & & &   \\
  2  & 3.64 & 3.96 & 4.35    \\
  5  & 3.70 & 4.25 & 4.63    \\
  10 & 3.85 & 4.35 & 4.76    \\
\tableline 
 $ T_c\ =\ 170\,$MeV & & &   \\
  2  & 3.40 & 3.71 & 4.05    \\
  5  & 3.60 & 4.00 & 4.35    \\
  10 & 3.76 & 4.20 & 4.50    \\
\tableline
 $ T_c\ =\ 200\,$MeV & & &   \\
  2  & 3.13 & 3.37 & 3.69    \\
  5  & 3.23 & 3.60 & 3.98    \\
  10 & 3.39 & 3.79 & 4.20    \\
\end{tabular}
\label{tab1}
\end{table}

\end{document}